\pdfoutput=1
\RequirePackage{ifpdf}
\ifpdf 
\documentclass[pdftex]{sigma}
\else
\documentclass{sigma}
\fi

\numberwithin{equation}{section}

\begin{document}


\renewcommand{\PaperNumber}{038}

\FirstPageHeading

\ShortArticleName{Mendeleev Table: a Proof of Madelung Rule and Atomic Tietz Potential}

\ArticleName{Mendeleev Table: a Proof of Madelung Rule\\ and Atomic Tietz Potential}

\Author{Eugene D.~BELOKOLOS}

\AuthorNameForHeading{E.D.~Belokolos}

\Address{Department of Theoretical Physics, Institute of Magnetism,\\
National Academy of Sciences of Ukraine, 36-b Vernadsky Blvd., Kyiv, 252142, Ukraine}
\Email{\href{mailto:bel@imag.kiev.ua}{bel@imag.kiev.ua}}

\ArticleDates{Received February 27, 2017, in f\/inal form May 22, 2017; Published online June 07, 2017}

\Abstract{We prove that a neutral atom in mean-f\/ield approximation has ${\rm O}(4)$ symmetry and this fact explains
the empirical $[n+l,n]$-rule or Madelung rule which describes ef\/fectively periods, structure
and other properties of the Mendeleev table of chemical elements.}


\Classification{81Q05; 81V45}

\section{Introduction}

In 1869 D.I.~Mendeleev discovered a periodic dependence of chemical element properties on $Z$ with periods
\begin{gather}2,\;8,\;8,\;18,\;18,\; 32,\; \ldots,\label{1}
\end{gather}
where $Z$ is a number of electrons in the chemical element atom\footnote{See \url{https://en.wikipedia.org/wiki/Periodic_table}.}. 

With creation of quantum mechanics physicists tried to explain the Mendeleev empirical law in terms of the one-particle quantum numbers
\begin{gather*}n,\;l,\;m,\; \sigma,
\end{gather*}
where $n=n_r+l+1$ is a principal quantum number, $n_r$ is a radial quantum number, $l$ is an orbital quantum number,
$m$ is a magnetic quantum number, $-l\leq m\leq l$, $\sigma=\pm 1/2$ is a~projection of electron spin.
For example, energy levels of the Hydrogen atom, which has ${\rm O}(4)$ symmetry, depend on the principal quantum number
$n$ only and are degenerate in other quantum numbers: $l$, $0\leq l\leq n-1$ (so called ``accidental'' degeneracy);
$m$, $-l\leq m\leq l$; $\sigma$, $\sigma=\pm 1/2$. It is a so called $[n,l]$-rule. According to it
the energy spectrum of atom consists of electron shells, enumerated by the principal quantum number $n$ and
having the degeneracy
\begin{gather*}N_n=\sum_{l=0}^{n}2(2l+1)=2n^2.
\end{gather*}
This formula gives the following set of periods
\begin{gather}2, \;8, \; 18, \; 32,\; \ldots.\label{2}
\end{gather}


Comparing the period sequences (\ref{1}) and (\ref{2}), we can explain only
the f\/irst two periods of the Mendeleev table. Although we see a remarkable fact
that (1) the period lengths have cardinalities that correspond to the Hydrogen degeneracy dimensions,
and (2) the same cardinalities always occur in pairs, except for the very f\/irst one.
In this paper we show that these similarities are not accidental.

\section[The Madelung $\protect{[n+l,n]}$-rule]{The Madelung $\boldsymbol{[n+l,n]}$-rule}

We get an exact expression for the Mendeleev periods and other properties of the Mendeleev table with the Madelung $[n+l, n]$-rule~\cite{Mad36}).

The $[n+l, n]$-rule asserts: with growth of atomic charge $Z$ the electrons f\/ill up in atom consecutively
the one-particle states with the least possible value of the quantum number $n + l$; and,
for a given value $n+l$, the electrons f\/ill up states with the least possible value of the quantum number~$n$.

Here it is reasonable to introduce the quantum number
\begin{gather*}M=n+l=n_r+2l+1.
\end{gather*}

The $[n+l,n]$-rule is in fact an algorithm for consecutive building-up of atoms\footnote{See \url{https://en.wikipedia.org/wiki/Aufbau_principle}.}. 
It describes the states for each of about 5000 electrons of the atoms to corresponding elements of the periodic system (indeed, now we have 118 chemical elements, and for approximately 100 elements we know every state of electron conf\/iguration, that is $\sum\limits_1^{100}Z=5050$ states) and predicts correctly the real electron conf\/igurations of all elements of periodic system with small number of exclusions.
There exist only~19 elements (Cr, Cu, Nb, Mo, Ru, Rh, Pd, Ag, La, Ce, Gd, Pt, Au, Ac, Th, Pa, U, Np, Cm), whose electron conf\/igurations dif\/fer from the conf\/igurations predicted with the $[n+l,n]$-rule~\cite{Meek02}.

The $[n+l,n]$-rule allows to obtain relations between the order number $Z$ of chemical element and quantum numbers $M$, $n$, $l$ of the appropriate electron conf\/iguration.

Let us designate the order number $Z$ of chemical element in which electrons from the $(n+l)$-subgroups, $(n, l)$-subgroups, $n$-subgroups, $l$-subgroups or $n_r$-subgroups appear for the f\/irst time by symbols
\begin{gather*}
 Z_{n+l},\; Z_{n, l},\; Z_n,\; Z_l,\; Z_{n_r}.
\end{gather*}
Let us designate an order number $Z$ of a chemical element in which the $(n+l)$-subgroups, $(n, l)$-subgroups, $n$-subgroups, $l$-subgroups or $n_r$-subgroups are f\/illed up completely for the f\/irst time by symbols
\begin{gather*}
 {}_{n+l}Z,\; {}_{n, l}Z,\; {}_nZ,\; {}_lZ,\; {}_{n_r}Z.
\end{gather*}

Then the $[n+l,n]$-rule with help of four actions of arithmetic and
well known formulas
\begin{gather*}\sum_{k=1}^{k=n}k = (1/2) n(n+1), \qquad \sum_{k=1}^{k=n}k^2 =
(1/6) n(n+1)(2n+1),
\end{gather*}
leads to the Klechkovski--Hakala formulas~\cite{Hak52,Klech68}
\begin{gather}
Z_{n+l}=K(n+l)+1,\qquad _{n+l}Z=K(n+l+1),\nonumber\\
Z_{n,l}=K(n+l+1)-2(l+1)^2+1,\qquad _{n,l}Z=K(n+l+1)-2l^2,\nonumber\\
Z_n=K(n+1)-1,\nonumber\\
{}_{n}Z=K(2n)-2(n-1)^2=(1/6)\big[(2n-1)^3+11(2n-1)\big],\nonumber\\
Z_l=K(2l+1)+1=(1/6)\big[(2l+1)^3 + (5-2l)\big],\nonumber\\
Z_{n_r}=K(n_{r}+2)-1.\label{KH}
\end{gather}
Here $K(x)$ is the following function
\begin{gather*}
K(x)=(1/6)x\big[x^2+2-3\mu(x)\big],\qquad x\in \mathbb{N},\\
\mu(x)=x\ \mbox{mod}\, (2)= \begin{cases}1,&x\ \mbox{is odd},\\0,& x\ \mbox{is even}.
 \end{cases}
\end{gather*}

Relations for $Z_{n+l}$ and $_{n+l}Z$ are exact. Other relations are exact up to exclusions pointed out above.
For example, for $l=0, 1, \ldots$, we have $Z_{l}=1, 5, 21, 57, \ldots$, that is,
according to the Madelung rule, the onset of the 4f block
starts with La ($Z=57$). But, according to the IUPAC data\footnote{See \url{https://iupac.org/what-we-do/periodic-table-of-elements/}.} on electron conf\/igurations in atoms, the onset of the 4f block starts with Ce ($Z=58$). 

The $[n+l, l]$-rule def\/ines essential characteristics of the Mendeleev periodic system.

In the Mendeleev periodic system, the table rows are enumerated by the Mendeleev num\-ber~$\textsf{M}$
which is a linear function of the Madelung number~$M$~\cite{Klech68}:
\begin{gather*}
\textsf{M}=M-1+\delta_{l,0}=n+l-1+\delta_{l,0},
\end{gather*}
where $\delta_{j,k}$ is the Kronecker delta.

It is easy to show that the number of elements in the Mendeleev $\textsf{M}$-th period
of the periodic system according to the $[n+l,n]$-rule is equal to
\begin{gather*}L_{\textsf{M}}=K(\textsf{M}+2)-K(\textsf{M}+1)
=2\left(\left[\frac{\textsf{M}}{2}\right]+1\right)^2,
\end{gather*}
where $[x]$ is an integer part of the real number $x$. Numbers $L_{\textsf{M}}$, $\textsf{M}=1,2,\ldots$ form the
sequence $2, 8, 8, 18, 18, 32, 32,\ldots, $ which coincides with empirical lengths~(\ref{1}) of the periods of
the system of elements.

According to the $[n+l,n]$-rule, the number $Z_{\textsf{M}}$ for the initial element
of the Mendeleev $\textsf{M}$-th period is equal to
\begin{gather*}
Z_{\textsf{M}}=K(\textsf{M}+1)-1,
\end{gather*}
and the number $_{\textsf{M}}Z$ for the f\/inal element of the Mendeleev $\textsf{M}$-th
period is equal to
\begin{gather*}
_{\textsf{M}}Z=K(\textsf{M}+2)-2.
\end{gather*}
The sequence $Z_{\textsf{M}}={1, 3, 11, 19, 37, 55, 87, \ldots}$ corresponds to alkaline metals, and
the sequence $_{\textsf{M}}Z={2, 10, 18, 36, 54, 86, \ldots}$ corresponds to noble gases.

Thus, the empirical $[n+l,n]$-rule is a very ef\/f\/icient method to explain periodic table and atomic properties.
Research to justify it continues up to now (see, e.g., \cite{Allen02,Kib04, Kib07, Ost81, Ost01}.
But after 80 years of studies we have not yet good understanding for it.
Standard textbooks on quantum mechanics even do not mention this rule.

In the present paper we give a theoretical basis for the $[n+l,n]$-rule. A preliminary version of this paper
was published in 2002 \cite{Bel02}.

Further we use everywhere the atomic units $\hbar=e=m=1$, where $\hbar$ is the reduced Plank constant,
$-e$ is the electron charge, and $m$ is the electron mass.

\section[Atomic potential in the mean-f\/ield and semi-classical approximations]{Atomic potential in the mean-f\/ield\\ and semi-classical approximations}

The Hamiltonian of a free neutral atom is
\begin{gather*}H_{\rm A} =\sum_{k=1}^Z \left(-\frac 12\Delta_k-
\frac{Z}{\vec{r}_k}\right)+\frac12\sum_{i,k=1}^Z\frac{1}{\vec{r}_{ik}}=
\sum_{k=1}^Z \left(-\frac12\Delta_k + v(\vec{r}	_k)\right),\\
v(\vec{r}_k) =-\frac{Z}{\vec{r}_k} + \frac12\sum_{i=1}^Z\frac{1}{\vec{r}_{ik}},
\end{gather*}
where $v(\vec{r})$ describes the electron interactions with the atomic nucleus and other electrons.
Since it is good approximation to describe the electrons in an atom in terms of the electron conf\/iguration,
i.e., the electron distribution on one-particle states,
\begin{gather*}
n, \ l, \ m, \ \sigma,
\end{gather*}
therefore we may consider the electron interaction $v(\vec{r})$ in the mean-f\/ield approximation
with the Hamiltonian
\begin{gather*}
 H=\sum_{k=1}^Z \left(-\frac12\Delta_k + V(\vec{r}_k)\right),
\end{gather*}
where $V(\vec{r})$ is the mean-f\/ield atomic potential. Since we consider a one-particle angular momentum, quantum number~$l$ as a good quantum number, this potential has to be central,
\begin{gather*}
V(\vec{r})=V(r).
\end{gather*}

For the Schr\"odinger equation
\begin{gather*}
 H\Psi=E\Psi
\end{gather*}
we look for the ground state solution in the Slater determinant form
\begin{gather*}
 \Psi=\det||\psi_j(r_k)||,
\end{gather*}
where one-particle wave function $\psi_k(r)$ satisf\/ies the one-particle Schr\"odinger equation
describing an electron in a 3-dimensional central potential $V(r)$,
\begin{gather*}
\left(-\frac12\Delta + V(r)\right)\psi_j(r)=E_j\psi_j(r).
\end{gather*}

In the semiclassical approximation, solutions of this equation have the form
\begin{gather*}
\psi(r)=\exp(iS(r)),
\end{gather*}
where $S(r)$ is the classical action.

\subsection{The Thomas--Fermi atomic potential}

Since in the case under consideration we have three integrals of
motion (energy $E$, angular momentum $L$, and its projection $L_z$) our problem is integrable
by quadratures due to the Arnold--Liouville theorem, and the radial action looks as follows
\begin{gather*}
 S(r)=\frac{1}{\pi}\int^r\left[2(E-V(r))-\frac{I_{\theta}^2}{r^2}\right]^{1/2}{\rm d}r,
\end{gather*}
where the radial action variable is
\begin{gather*}
I_r=\frac{1}{\pi}\int_{r_{\min}}^{r_{\max}}\left[2(E-V(r))-\frac{I_{\theta}^2}{r^2}\right]^{1/2}{\rm d}r.
\end{gather*}
In the semiclassical approximation we should change action variables by quantum numbers,
\begin{gather*}I_r= n_r(E,l),\qquad I_{\theta}= l.
\end{gather*}
In this way we come to the Bohr--Sommerfeld quantization rule
\begin{gather*}n_r(E,l)=\frac{1}{\pi}\int_{r_{-}(E,l)}^{r_{+}(E,l)}
\left[2E-2V(r)-\frac{l^2}{r^2}\right]^{1/2}\mathrm{d}r.
\end{gather*}

Let us set in the latter expression $E=0$, then
\begin{gather*} n_r(0,l)=\frac{\sqrt{2}}{\pi}\int_{r_{-}(0,l)}^{r_{+}(0,l)}
\left[-V(r)-\frac{l^2}{2r^2}\right]^{1/2}\mathrm{d}r.
\end{gather*}
Thus the number of bound states $N$ in the atom is
\begin{gather*}N=\sum_{l=0}^{l^{\max}}n_r(0,l)2(2l+1).
\end{gather*}
Approximately a value $N$ looks as follows
\begin{gather*} N\simeq \int_{l=0}^{l^{\max}} n_r(0,l)2(2l+1)\mathrm{d}l\\
\hphantom{N}{} =\frac{\sqrt{2}}{\pi}\int_{r_{-}}^{r_{+}}\int_{l=0}^{l^{\max}}
\left[-V(r)-\frac{l^2}{2r^2}\right]^{1/2}2(2l+1)\mathrm{d}l\mathrm{d}r\\
\hphantom{N}{}=\frac{\sqrt{2}}{\pi}\frac{2}{3}\int_{r_{-}}^{r_{+}}
\left[-V(r)\right]^{3/2}4r^2\mathrm{d}r
=\frac{2\sqrt{2}}{3\pi^2}\int_{r_{-}}^{r_{+}}
\left[-V(r)\right]^{3/2}4\pi r^2\mathrm{d}r =\int_{r_{-}}^{r_{+}}\rho(r)4\pi r^2\mathrm{d}r.
\end{gather*}
This is the well-known asymptotic formula \cite{Reed1978} for the number of bound states in a~central potential
with the density of bound states $\rho(r)$ equal to
\begin{gather*}\rho(r)=\frac{2\sqrt{2}}{3\pi^2}[-V(r)]^{3/2}.
\end{gather*}
The electrostatic potential of atomic electrons $\phi(r)=-V(r)$ satisf\/ies the Poisson equation $\Delta\phi(r)=-4 \pi\rho$,
and therefore
\begin{gather*} \Delta\phi(r)=\frac{8\sqrt{2}}{3\pi}\phi^{3/2}(r),\\
 \phi(r)r=Z,\qquad r\to 0,\qquad \phi(r)=0,\qquad r\to\infty.
\end{gather*}
We can present the Thomas--Fermi potential $\phi(r)$ in such a way
\begin{gather*}\phi(r)=\frac{Z}{r}\chi(x),\qquad x=\frac{r}{R_0},\qquad R_0=bZ^{-1/3},
\qquad b=\frac{1}{2}\left(\frac{3\pi}{4}\right)^{2/3}\simeq 0.885.
\end{gather*}

For large $Z$ the ground state energy $E_{\rm TF}(Z)$ of the Thomas--Fermi atom is the asymptotics for
a ground state energy $E_{\rm HF}(Z)$ of the Hartree--Fock atom \cite{Lieb81, Lieb77}:
\begin{gather*}
 \lim_{Z\to\infty}E_{\rm HF}(Z)/E_{\rm TF}(Z)=1.
\end{gather*}

The Thomas--Fermi potential does not depend on quantum numbers although naturally it should do. Nevertheless,
let us calculate the $Z_l$ for the Thomas--Fermi potential \cite{Fermi28}.

The ef\/fective potential $u_l(r)$ is
\begin{gather*}u_l(r) =-\phi(r)+\frac{(l+1/2)^2}{2r^2}=
-\frac{(l+1/2)^2}{2r^2} [ \zeta_lx\chi(x)-1 ],
\end{gather*}
where
\begin{gather*}
\zeta_l =\frac{2ZR_0}{(l+1/2)^2}=2b\frac{Z^{2/3}}{(l+1/2)^2}.
\end{gather*}

Conditions for the f\/irst appearance of the energy level with certain $l$ are
\begin{gather*}u_l(r)=0,\qquad u_l^{\prime}(r)=0
\end{gather*}
or, which is the same,
\begin{gather*}u_l(x)=0,\qquad u_l^{\prime}(x)=0.
\end{gather*}
It is equivalent to equations
\begin{gather*}\zeta_l x\chi(x)-1=0,\qquad \frac{\mathrm{d}}{\mathrm{d}x}(x\chi(x))=0,
\end{gather*}
which have a solution
\begin{gather*} x_0\simeq 2.104,\qquad \chi(x_0)\simeq 0.231,\qquad x_0\chi(x_0)\simeq 0.486,\\
\chi'(x_0)\simeq-0.110,\qquad \zeta_l=[x_0\chi(x_0)]^{-1}\simeq 2.056.
\end{gather*}
This means that
\begin{gather*}\frac{Z_l^{\rm TF}}{(2l+1)^3}=\left(\frac{\zeta_l}{8b}\right)^{3/2}=0.155,
\end{gather*}
and hence
\begin{gather*}Z_l^{\rm TF}\simeq 0.155(2l+1)^3.
\end{gather*}
In the general case $Z_l^{TF}$ is not integer and we have to write
\begin{gather*}Z_l^{\rm TF}= \big|0.155(2l+1)^3\big|,
\end{gather*}
where $|x|$ means the integer which is the closest to the real $x$.

If we change in this formula the coef\/f\/icient $0.155$ to $0.169\simeq 1/6$ we get a~better agreement of
the values $Z_l^{TF}$ with that in the Mendeleev table (see \cite{Iwan53} and \cite[Section~73]{LL74}). In this case
the r.h.s.\ of the formula will coincide with the f\/irst summand of the Klechkovski--Hakala expression
for $Z_l$~(\ref{KH}) obtained with the $[n+l]$-rule.

This example shows that the Thomas--Fermi theory yields only some approximation to the $[n+l,n]$-rule.

According to the above discussion the electrons in the atom in the mean-f\/ield approximation interact by means of
central atomic potential, having geometrical ${\rm O}(3)$ symmetry. The Thomas--Fermi theory takes into account
the ${\rm O}(3)$ symmetry of the atomic Hamiltonian with central potential. We do the next step.
The Hamiltonian with any central potential
has dynamical~${\rm O}(4)$ symmet\-ry~\cite{Bacry66, Fradkin67, Mukunda67}, similar to the Hydrogen
atom~\cite{Barg36, Fock35, Pauli26}. Therefore, the atomic Hamiltonian must have an additional
integral of motion in involution with respect to integrals of motion corresponding to~${\rm O}(3)$ symmetry.
Further we shall show that such additional integral of motion in involution does exist for the atomic Hamiltonian
and leads to $[n+l,n]$-rule.

\subsection{The Tietz atomic potential}

We begin with certain basic facts on symmetry and integrability in the classical Hamiltonian systems
(see \cite[Sections~49--51]{Arn89}, \cite[Section~52]{LL73} and~\cite{Tor16}).

Let us assume that on a symplectic $2n$-dimensional manifold there are $n$ functions in involution
\begin{gather*}
 F_1,\dots, F_n,\qquad \{F_i,F_j\}=0,\qquad i,j= 1, \dots, n.
\end{gather*}
Consider a level set of functions $F_i$
\begin{gather*}
 M_f=\{x\colon F_i(x)=f_i,\, i=1,\dots, n \}.
\end{gather*}
Suppose that $n$ functions~$F_i$ are independent on $M_f$ (i.e., the $n$ $1$-forms $dF_i$ are linearly independent in every point of~$M_f$).

Then
\begin{enumerate}\itemsep=0pt
\item $M_f$ is a smooth manifold, invariant with respect to the phase f\/low with the Hamiltonian $H=F_1$.

\item If the manifold $M_f$ is compact and connected, then it is dif\/feomorphic to $n$-dimensional torus
\begin{gather*}
 T^n=\{(\phi_1,\dots,\phi_n) \ (\mbox{mod}\ 2 \pi)\}.
\end{gather*}

\item Phase f\/low with the Hamiltonian $H$ def\/ines on $M_f$ a quasi-periodic movement
with angular variables $(\phi_1,\dots,\phi_n)$,
\begin{gather*}
 \frac{{\rm d}\phi_i}{{\rm d}t}=\omega_i(f),\qquad \phi_i(t)=\phi_i(0)+\omega_i t,\qquad i=1,\dots, n.
\end{gather*}
Instead of functions $F=(F_1,\dots, F_n)$ it is possible to def\/ine new functions
$I=(I_1(F),\dots$, $I_n(F))$ which are called action variables and which together with angle variables
form in the neighborhood of manifold $M_f$ the canonical system of action-angle coordinates.

\item Canonical equations with the Hamiltonian function are integrable in quadratures.

\item If frequencies $\omega=(\omega_1,\dots, \omega_n)$
are degenerated, i.e., if there exists such an integer-valued vector $k=(k_1,\dots, k_n)\in\mathbb{Z}^n$ that
\begin{gather*}
 (k,\omega)=0,
\end{gather*}
then there appears one more single-valued function $F_{n+1}$ which is in involution with functions $F_1,\dots, F_n$.
\end{enumerate}

Now let us go back to our problem: an electron in the central atomic potential. In this case the 3-dimensional electron movement
is reduced to 2-dimensional one in the plane perpendicular to the orbital momentum vector. We describe this movement
in the semiclassical approximation by the action
\begin{gather*}
S=\int^{r}\sqrt{2 [E-V(r)]-\frac{I_{\theta}^2}{r^2}} \mathrm{d}r
\end{gather*}
with radial $(r, I_r)$ and orbital $(\theta, I_{\theta})$ action-angle variables, where
\begin{gather*}
I_r=\frac{1}{\pi}\int^{r_{\max}}_{r_{\min}}\sqrt{2 [E-V(r) ]-\frac{I_{\theta}^2}{r^2}} \mathrm{d}r.
\end{gather*}
And so in our problem we have a quasi-periodic movement on 2-dimensional torus which is described by the Fourier series
\begin{gather*}
G(t)=\sum_{l_1\in\mathbb{Z}}\sum_{l_2\in\mathbb{Z}}G_{l_1,l_2}\exp(l_1\phi_r+l_2\phi_{\theta})
=\sum_{l_1\in\mathbb{Z}}\sum_{l_2\in\mathbb{Z}}G_{l_1,l_2}\exp[(l_1\omega_r+l_2\omega_{\theta})t]
\end{gather*}
with 2 basic frequencies
\begin{gather*}
\omega_r=\frac{\partial E}{\partial I_r},\qquad \omega_{\theta}=\frac{\partial E}{\partial I_{\theta}}.
\end{gather*}
In the general case these frequencies are independent. But under certain circumstances, as we have,
when new additional integral appears, they may become degenerate (or commensurate, or resonance),
\begin{gather*}q\omega_r=p\omega_{\theta},\qquad q, p \in\mathbb{Z},\qquad {\rm g.c.d.}(p, q)=1,
\end{gather*}
where ${\rm g.c.d.}(p,q)$ means the greatest common divisor of the integers $p$ and $q$.
This condition leads to important consequences.
\begin{enumerate}\itemsep=0pt
\item Since the latter relation is equivalent to
\begin{gather*}
q\frac{\partial E}{\partial I_r}=p\frac{\partial E}{\partial I_{\theta}}
\end{gather*}
it means that energy $E$ depends on $pI_r+qI_{\theta}$:
\begin{gather*}E=E(pI_r+qI_{\theta}).
\end{gather*}
And due to the Bohr--Sommerfeld semiclassical quantization rule
\begin{gather*}I_r\rightarrow n_r,\qquad I_{\theta}\rightarrow l,
\end{gather*}
we have
\begin{gather*}E=E(p n_r+q l).
\end{gather*}

\item We have the additional independent integral of motion.

\item The canonical variables of the problem are separated in several systems of coordinates.
\end{enumerate}

For example, for the Kepler--Coulomb potential $1/r$ we have $p=q=1$, energy $E$ depends on
principal quantum number $n=n_r +l +1$, the canonical variables of the problem are separated in polar and parabolic
coordinates.

Let us study the similar situation for the atomic potential. Consider a system of equations
\begin{gather*}
I_r(E) =\frac{1}{\pi}\int_{r_{-}(E)}^{r_{+}(E)}
\left[2E-2V(r)-\frac{I_{\theta}^2}{r^2}\right]^{1/2}\mathrm{d}r,\\
M_{p,q} = pI_r+qI_{\theta},\qquad q, p \in\mathbb{Z},\qquad {\rm g.c.d.}(p, q)=1,
\end{gather*}
the f\/irst of which is the standard relation between action variables $I_r$, $I_{\theta}$, and energy $E$
in central potential and the second one is relation between $I_r$, $I_{\theta}$ arising as result of frequencies
degeneracy.

We shall study equations at the energy $E=0$ at which new bound states appear from continuous spectrum, when
$Z$ grows. Then dif\/ferentiating equations with respect to $I_{\theta}$ we come to the equation
\begin{gather*}\pi\alpha=\int_{r_{-}}^{r_{+}}\big[-2V(r)r^2-I_{\theta}^2\big]^{-1/2}
I_{\theta}\frac{\mathrm{d}r}{r},
\end{gather*}
where
\begin{gather*}
\alpha=-\frac{\partial I_r}{\partial I_{\theta}}=\frac{q}{p}\in\mathbb{Q},\qquad q, p \in\mathbb{Z},
\qquad {\rm g.c.d.}(p, q)=1.
\end{gather*}

From this equation we deduce an expression for atomic potential $V(r)$. A~similar integral equation arising
in problem for generalized tautochrone (isochrone) curve was solved
by Abel~\cite{Abel1823}. We shall use his method in the form presented in \cite[Section~12]{LL73}.

\begin{theorem}
Equation
\begin{gather}\label{eq0}
\pi\alpha=\int_{r_{-}}^{r_{+}}\big[{-}2V(r)r^2-L^2\big]^{-1/2}
L\frac{\mathrm{d}r}{r}
\end{gather}
has a solution
\begin{gather*}
V_{\alpha, \beta, R}(r)=-\frac{\beta}{r^2[(r/R)^{1/\alpha}+(R/r)^{1/\alpha}]^2},
\end{gather*}
where $\beta$ and $R$ are certain constants, and its deformations.

If the potential $V_{\alpha, \beta, R}(r)$ coincides at small $r$ with the Kepler--Coulomb potential
\begin{gather*}V_{\alpha, \beta, R}=-Z/r,\qquad r\to 0,
\end{gather*}
then we have
\begin{gather*}\alpha=q/p=2,\qquad \beta=ZR,
\end{gather*}
and the potential takes the following form
\begin{gather*}V(r) \equiv V_{2,ZR,R}(r)=-\frac{Z}{r(1+(r/R))^2}.
\end{gather*}
\end{theorem}

\begin{remark}
According to the theorem in the neighborhood $E\simeq 0$ we have
\begin{gather*}E=E(n+l)
\end{gather*}
and this fact proves the f\/irst part of the $[n+l, n]$-rule.
A second part of the $[n+l, n]$-rule is a~consequence of the oscillation theorem.
\end{remark}

\begin{proof}
Let us rewrite equation (\ref{eq0}) in the form
\begin{gather*}\pi\alpha=\int_{r_{-}}^{r_{+}}\big[w(x)-L^2\big]^{-1/2}L\mathrm{d}x,
\end{gather*}
where
\begin{gather*} x=\ln(r/R),\qquad r=R\exp(x),\\
w(x)=-2V(r)r^2 \qquad \mbox{at}\quad r=R\exp(x),\\
w_0=\max w(x),
\end{gather*}
are new variables and $R$ is a parameter.
We assume that $w(x)$ is the one-well potential
and therefore the inverse function is two-valued,
i.e., the values $w$ are reached in two points $x_{-}(w)$ and $ x_{+}(w)$. We shall assume also that
$x_{-}(w)\leq x_{+}(w)$ and at $w_0$ we have $x_{-}(w_0)=x_{+}(w_0)$. As a result we obtain
\begin{gather*}\pi\alpha=\int_{L^2}^{w_0}\big(w-L^2\big)^{-1/2}\left(\frac{\mathrm{d}x_{+}}{\mathrm{d}w}
-\frac{\mathrm{d}x_{-}}{\mathrm{d}w}\right)L\mathrm{d}w.
\end{gather*}

Multiplying this equality by $\big(L^2-w_{1}\big)^{-1/2} 2\mathrm{d}L$ and integrating from $(w_{1})^{1/2}$
to $(w_{0})^{1/2}$ we get
\begin{gather*} 2\pi\alpha\int_{(w_1)^{1/2}}^{(w_0)^{1/2}}\big(L^2-w_1\big)^{-1/2}\mathrm{d}L\\
\qquad{} =\int_{(w_1)^{1/2}}^{(w_0)^{1/2}}2L\mathrm{d}L\int_{L^2}^{w_0}\left(\frac{\mathrm{d}x_{+}}{\mathrm{d}w}
-\frac{\mathrm{d}x_{-}}{\mathrm{d}w}\right)\big[\big(w-L^2\big)\big(L^2-w_{1}\big)\big]^{-1/2}\mathrm{d}w\\
\qquad {} =\int_{w_1}^{w_0}\mathrm{d}w\left(\frac{\mathrm{d}x_{+}}{\mathrm{d}w}
-\frac{\mathrm{d}x_{-}}{\mathrm{d}w}\right)
\int_{(w_1)^{1/2}}^{(w)^{1/2}}\big[\big(w^2-L^2\big)\big(L^2-w_1\big)\big]^{-1/2}2L\mathrm{d}L.
\end{gather*}
Since
\begin{gather*} \int_{(w_1)^{1/2}}^{(w_0)^{1/2}}\big(L^2-w_1\big)^{-1/2}\mathrm{d}L=\operatorname{Arcosh}(w_0/w_1)^{1/2},\\
\int_{(w_1)^{1/2}}^{(w)^{1/2}}\big[\big(w^2-L^2\big)\big(L^2-w_1\big)\big]^{-1/2}2L\mathrm{d}L=\pi,
\end{gather*}
this equality acquires the following form
\begin{gather*}2\alpha \operatorname{Arcosh}(w_0/w_1)^{1/2}=(x_{+}(w)-x_{-}(w))\Big|_{w_1}^{w_0}.
\end{gather*}
Taking into account that $x_{-}(w_0)=x_{+}(w_0)$ and setting $w_{1}=w$ we come to the equality
\begin{gather*}x_{+}(w)-x_{-}(w)=2\alpha \operatorname{Arcosh}(w_0/w)^{1/2}.
\end{gather*}

This equality def\/ines only the dif\/ference $x_{-}(w)-x_{+}(w)$ of two functions $x_{-}(w)$ and $ x_{+}(w)$, any
of which remains actually undef\/ined. It means that there exists the inf\/inite set of potentials which satisfy
the latter equation and dif\/fer by deformations which do not change the dif\/ference of two values
of $x$ corresponding to one value of~$w$. Among these potentials there is a symmetric potential with the property
$x_{+}(w)=-x_{-}(w)\equiv x(w)$. For this potential in this case
\begin{gather*}w(x)=w_{0}/\cosh^2(x/\alpha),
\end{gather*}
or, in the previous notations,
\begin{gather*}V_{\alpha, \beta, R}(r)=-\frac{\beta}{r^2[(r/R)^{1/\alpha}+(R/r)^{1/\alpha}]^2},
\end{gather*}
where $\beta=2w_{0}$. These potentials are degenerate for arbitrary value of the parameter $\alpha$.
In order for the degenerate potential to coincide with Coulomb potential at small $r$
\begin{gather*}V_{\alpha, \beta, R}=-Z/r,\qquad r\to 0,
\end{gather*}
we must have
\begin{gather*}\alpha=q/p=2,\qquad \beta=ZR.
\end{gather*}
Then this potential takes the following form
\begin{gather*}V(r) \equiv V_{2,ZR,R}(r)=-\frac{Z}{r(1+(r/R))^2}.\tag*{\qed}
\renewcommand{\qed}{}
\end{gather*}\renewcommand{\qed}{}
\end{proof}

In this potential the total number of bound states $N$ is equal to the total number of elect\-rons~$Z$ if
\begin{gather*}N=\frac{8\sqrt{2}}{3\pi}\int[-V(r)]^{3/2}{\rm d}r=\big(9/2R^3\big) =Z,
\end{gather*}
i.e.,
\begin{gather*}R=(9/2Z)^{1/3}\simeq 1.651 Z^{-1/3},
\end{gather*}
and we have f\/inally
\begin{gather*}V(r|Z) =-\frac{Z}{r(1+(r/R(Z)))^2}.
\end{gather*}
For the given electron conf\/iguration we must use the Klechkovski--Hakala
formulas for~$Z$. In this case the potential will depend on quantum numbers.

We can present this potential as the sum
\begin{gather*}V(x)=-\frac{(Z/R)}{x(1+x)^2}=-\frac{Z}{R}\left[\frac{1}{x}-\frac{1}{(x+1)}
-\frac{1}{(x+1)^2}\right], \qquad x=\frac{r}{R},
\end{gather*}
where the f\/irst summand describes the Coulomb attraction of the atomic nucleus for the single electron and
two other summands describe a nucleus screening by the other electrons.

T.~Tietz \cite{Tietz55} proposed the potential
\begin{gather*} V(r) = - \frac{Z}{r[1 + (r/R)]^2}
\end{gather*}
as a good rational approximation to the Thomas--Fermi potential and used it for calculation of various atomic
properties and explanation of the periodic system elements~\cite{Tietz60a} (see also~\cite{Wong79}). Further we
shall call this potential as the Tietz atomic potential.

Due to the proximity of the Tietz and Thomas--Fermi potentials it is very likely that at large~$Z$
a~ground state energy of the Tietz atom, $E_{\rm T}(Z)$, is an asymptotics for
the ground state energy of the Hartree--Fock atom, $E_{\rm HF}(Z)$:
\begin{gather*}
 \lim_{Z\to\infty}E_{\rm HF}(Z)/E_{\rm T}(Z)=1.
\end{gather*}

Yu.N.~Demkov and V.N.~Ostrovski \cite{Dem71} pointed out that this potential is a particular case of a so called
focussing (in other words ``degenerate'') potentials studied in connection to certain problems of optics
by J.C.~Maxwell \cite{Maxwel52} and V.~Lenz~\cite{Lenz28}.

The Tietz potential is a rational function, and thus we can easily do various calculations with it. For example, we can calculate the atomic spectrum.

\section{Semiclassical atomic spectrum for the Tietz atomic potential}

The Bohr--Sommerfeld semiclassical condition of quantization for the Tietz atomic potential is
\begin{gather*}
n_r=\frac{1}{\pi}\int_{r_-}^{r_+}\left[2E+\frac{2Z}{r(1+(r/R))^2}-\frac{(l+1/2)^2}{r^2}\right]^{1/2}{\rm d}r.
\end{gather*}
In the scaled quantum numbers
\begin{gather*}
x=\frac{r}{R},\qquad\epsilon=\frac{2ER^2}{(l+1/2)^2},\qquad \nu_r=\frac{n_r}{l+1/2},\qquad \eta_l=\frac{2ZR}{(l+1/2)^2}
\end{gather*}
the Bohr--Sommerfeld equation takes the form
\begin{gather*}
\nu_r=\frac{1}{\pi}\int_{x_-}^{x_+}\left[\epsilon+\frac{\eta_l}{x(1+x)^2}-\frac{1}{x^2}\right]^{1/2}{\rm d}x
=\frac{1}{\pi}\int_{x_-}^{x_+}\frac{\sqrt{P_4(x)}}{x(1+x)}{\rm d}x,
\end{gather*}
where $P_4(x)$ is a 4-th degree polynomial
\begin{gather*}
P_4(x)=\epsilon x^2(1+x)^2+\eta_l x-(1+x)^2,
\end{gather*}
and boundaries of integration are real non-negative zeros of this polynomial.
Therefore scaled radial quantum number $\nu_r$ is a period of the elliptic integral
with scaled energy $\epsilon$ and scaled charge~$\eta_l$ as the parameters.
We can obtain an atomic spectrum in the semiclassical approximation
\begin{gather*} \epsilon = f (\nu_r,\eta_l ),
\end{gather*}
by means of inversion of the elliptic integral.

Let us study this problem in the framework of perturbation theory in the energy $\epsilon$.
We present the Bohr--Sommerfeld semiclassical equation in the form
\begin{gather*}
\nu_r =\frac{1}{\pi}\int_{x_-}^{x_+}\left[\epsilon +\frac{\eta_l x-(1+x)^2}{x^2(1+x)^2}\right]^{1/2}{\rm d}x \\
\hphantom{\nu_r}{} =\frac{1}{\pi}\int_{x_-}^{x_+}\left[\epsilon +\frac{(x_+ -x)(x-x_-)}{x^2(1+x)^2}\right]^{1/2}{\rm d}x,
\end{gather*}
where
\begin{gather*}
\eta_l x-(1+x)^2=(x_+ -x)(x-x_-),\\
x_{\pm}^2-(\eta_l-2)x_{\pm} +1=0,\\
x_{\pm}=\frac{(\eta_l-2)\pm\sqrt{(\eta_l-2)^2-4}}{2}=
\frac{(\eta_l-2)\pm\sqrt{\eta_l(\eta_l-4)}}{2},\\
-(\eta_l-4)>\epsilon>-\frac{1}{4}\eta_l(\eta_l-4).
\end{gather*}

At the f\/irst approximation we have
\begin{gather*}
\nu_r\simeq J_0+\epsilon J_1 \\
\hphantom{\nu_r}{} = \frac{1}{\pi}\int_{x_-}^{x_+}[(x_+ -x)(x-x_-)]^{1/2}\frac{{\rm d} x}{x(1+x)}
+\epsilon \frac{1}{2\pi}\int_{x_-}^{x_+}
\frac{x(1+x)}{[(x_+ -x)(x-x_-)]^{1/2}}{\rm d} x,
\end{gather*}
where
\begin{gather*}
J_0=\frac{1}{\pi}\int_{x_-}^{x_+}[(x_+ -x)(x-x_-)]^{1/2}\frac{{\rm d} x}{x(1+x)}=\sqrt{\eta_l}-2,\\
J_1=\frac{1}{2\pi}\int_{x_-}^{x_+}\frac{x(1+x)}{[(x_+ -x)(x-x_-)]^{1/2}}{\rm d} x
=\frac{1}{16}\big(3\eta_l^2-8\eta_l\big).
\end{gather*}

The energy spectrum in the semiclassical theory is{\samepage
\begin{gather*}
\epsilon_{n,l} =-\frac{J_0-\nu_r}{J_1}=-16\frac{\sqrt{\eta_l}-2-\nu_r}{\big(3\eta_l^2-8\eta_l\big)}
=-16\frac{(l+(1/2))\sqrt{\eta_l}-2l-1-n_r}{(l+(1/2))\big(3\eta_l^2-8\eta_l\big)}\\
\hphantom{\epsilon_{n,l}}{} =-16\frac{(l+(1/2))\sqrt{\eta_l}-(n+l)}{(l+(1/2))\big(3\eta_l^2-8\eta_l\big)}
=-16\frac{\sqrt{2ZR}-M}{(l+(1/2))\big(3\eta_l^2-8\eta_l\big)},
\end{gather*}
where we have used the Madelung number $M=n+l$.}

Since
\begin{gather*}\epsilon_{M,l}=\frac{2E_{M,l}R^2}{(l+(1/2))^2},
\end{gather*}
we have the following expression for the atomic spectrum in the atomic units
\begin{gather*}E_{M,l}
=-8\frac{(\sqrt{2ZR}-M)(l+(1/2))}{\big(3\eta_l^2-8\eta_l\big)R^2},\qquad
\eta_l=\frac{2ZR}{(l+(1/2))^2}.
\end{gather*}

If
\begin{gather*}
\sqrt{2ZR}=M
\end{gather*}
then $E_{M,l}=0$, and all states $(M,l)$ are degenerate with respect to~$l$. Thus at energy $E=0$ there appears the full set of energy levels $(n, l)$ with $M=n+l$. Since $R=(9/2Z)^{1/3}$ the equality $M=\sqrt{2ZR}$ is equivalent to
\begin{gather*}
Z=(1/6)M^3.
\end{gather*}

We can write down a complete perturbation series
\begin{gather}\label{ser}
\nu_r=\sum_{k=0}^{\infty}\epsilon^k J_k.
\end{gather}
Since all integrals
\begin{gather*} J_k=\frac{c_k}{\pi}\int_{x_-}^{x_+}
\left\{\frac{x^2(1+x)^2}{[(x_+ -x)(x-x_-)]}\right\}^{k-(1/2)}\mathrm{d}x,\\
c_k=\frac{\Gamma(3/2)}{\Gamma(k+1)\Gamma((3/2)-k)},\qquad k\geq 2,
\end{gather*}
are divergent we should take their regularized values. i.e., valeur principale.
Inverting series (\ref{ser}) by means of the B\"{u}rmann--Lagrange theorem
we can obtain an expression for atomic spectrum in the semiclassical approximation
\begin{gather*} \epsilon = f (\nu_r,\eta_l ).
\end{gather*}

We plans to compare this atomic spectrum (both eigenvalues and eigenfunctions) with those
presented in articles \cite{Latter55,Tietz60b}.

\section{Conclusion}

For the Mendeleev periodic system of elements we have proved the empirical $(n + l,n)$-rule
which explains very ef\/f\/iciently the structure and properties of chemical elements.
In order to prove this rule we are forced to build for the Hamiltonian describing an electron in
central atomic potential one more integral of motion in involution in addition to energy and
orbital integral. In this case the atomic potential appears to be the Tietz potential.
For the Tietz potential we have calculated the atomic energy spectrum in the semiclassical approximation.

In addition, we are going to study spectrum of the Schr\"odinger operator with the Tietz potential,
when the radial part of wave function satisf\/ies the conf\/luent Heun equation
and compare energy levels, obtained in such a way, with the NIST Atomic Spectra Database.

For the atom in the paper we have studied the non-relativistic atomic Hamiltonian. We plan to consider
also the relativistic Dirac Hamiltonian.

It is interesting also to study isospectral many-well deformation of the Tietz potential.

Questions studied in the paper are related only to the ground state of atoms. But according to
experimental data the weakly excited atomic states also follow to $[n+l, n]$-rule \cite{Klech68}. It may be
of interest for physical chemistry, especially for understanding chemical reactions rules.

We hope to do that in the near future.

\pdfbookmark[1]{References}{ref}
\LastPageEnding

\end{document}